# A PROBLEM OF HYPOTHETICAL EMERGING OF MATTER OBJECTS ON HORIZON IN THE STANDARD MODEL OF UNIVERSE

**Vladimír Skalský**

Faculty of Materials Science and Technology of the Slovak Technical University, 917 24 Trnava, Slovakia, skalsky@mtf.stuba.sk

**Abstract.** In the standard model of universe the increase in mass of our observed expansive Universe is explained by the assumption of emerging the matter objects on the horizon (of the most remote visibility). However, the physical analysis of the influence of this assumption on the velocity of matter objects shows unambiguously that this hypothetical assumption contradicts the theory of gravity.

From *the Planck quantum hypothesis* (Planck 1899) and *the dimensional analysis* it results that our observed *expansive and isotropic relativistic Universe* started its expansive evolution with these beginning parameters:

*the Planck mass*

$$m_{Planck} = \sqrt{\frac{\hbar c}{G}} \sim 10^{-8} \text{ kg}, \qquad (1)$$

*the Planck length*

$$l_{Planck} = \sqrt{\frac{\hbar G}{c^3}} \sim 10^{-35} \text{ m}, \qquad (2)$$

*the Planck time*

$$t_{Planck} = \sqrt{\frac{\hbar G}{c^5}} \sim 10^{-43} \text{ s}, \qquad (3)$$

*the Planck mass density*

$$\rho_{Planck} = \frac{c^5}{\hbar G^2} \sim 10^{97} \text{ kg m}^{-3}. \qquad (4)$$

At present time these *present parameters of the Universe* are estimated:

*the present mass (of "the visible Universe")*

$$m_{pres} \sim 10^{53} \text{ kg}, \qquad (5)$$

*the present gauge factor*

$$a_{pres} \sim 10^{26} \text{ m}, \qquad (6)$$

*the present cosmological time*

$$t_{pres} \sim 10^{10} \text{ yr}, \qquad (7)$$

*the present mass density*

$$\rho_{pres} \sim 10^{-26} \text{ kg m}^{-3}. \qquad (8)$$

*Note*: In the available cosmological literature the value of $\rho_{pres}$ is estimated in the extent: $\rho_{pres} \sim 10^{-26} - 10^{-32}$ kg m$^{-3}$. In the relation (8) we reduced $\rho_{pres}$ to the largest estimated value, because other values mathematically do not correspond to the relations $m_{pres}$ (5) and $a_{pres}$ (6).

According to the relations (2) and (3), the Universe began its expansive evolution at the velocity

$$v = c, \qquad (9)$$

where $c$ is the boundary velocity of signal propagation.

When comparing the relations (6) and (7) it results the relation for the gauge factor of the Universe $a$ and the cosmological time of the Universe $t$:

$$a \sim ct. \qquad (10)$$

From comparison of the relations (1) and (5); (2) and (6); and (3) and (7) it results that the mass of Universe *m*, the gauge factor of Universe *a* and the cosmological time of Universe *t* during expansive evolution of our Universe increased approximately by 60 ranges.

In *the standard model of universe* the increase in mass of our Universe is explained by *the hypothetical assumption of emerging matter objects* on *the horizon (of the most remote visibility, the horizon of events)*.

From comparison of the relations (1), (2), (3), (4), (5), (6), (7) and (8) it results that the mass of Universe *m* increases proportionally to *a*, the volume of Universe *V* grows proportionally to $a^3$ and the mass density of Universe $\rho$ decreases proportionally to $a^{-2}$. Therefore, the hypothetical matter objects emerging on the Universe horizon (of the most remote visibility) would have to be demonstrated by the rapid slowing down of their expansion. It can be verified by a simple calculation:

From the relation for the mass of homogeneous matter sphere *m* with the radius *r* and the mass density $\rho$:

$$m = \frac{4}{3}\pi r^3 \rho \tag{11}$$

it results that the chosen starting mass of universe $m_0$ in the expansive homogeneous universe – the mass of which increases as a result of emergence of matter objects on the horizon (of the most remote visibility) – could be extended to the volume of homogeneous matter sphere with the resulting radius $r_x$ and the resulting universe mass density $\rho_x$:

$$r_x = \sqrt[3]{\frac{3m_0}{4\pi\rho_x}} . \tag{12}$$

The matter objects in the expansive homogeneous universe – in which the matter objects emerge on the horizon (of the most remote visibility) – and, according to the relation (10), expand at the velocity $v \sim c$, when they appear in the distance $r_x$, expand at the velocity

$$v_{r(x)} \sim \frac{r_x}{a_x} c , \tag{13}$$

where $a_x$ is the corresponding resulting gauge factor.

From the relations (1), (8) and (12) it results that the mass of "Planckton" $m_{Planck}$ (1) at the present cosmological time $t_{pres}$ (7) could be extended to the volume of homogeneous matter sphere with the present mass density $\rho_{pres}$ (8) and the radius (Skalský 1996, 1997)

$$r_{Planck-pres} = \sqrt[3]{\frac{3m_{Planck}}{4\pi\rho_{pres}}} \sim 8 \times 10^5 \text{ m}. \tag{14}$$

From the relations (6), (13) and (14) it results that the velocity of the initial "Planckton" $v = c$ in the homogeneous Universe with the present gauge factor $a_{pres}$ (6) in the distance of a homogeneous matter sphere with the radius $r_{Planck-pres}$ (14) it must decrease to the velocity (Skalský 1996, 1997)

$$v_{r(Planck-pres)} \sim \frac{r_{Planck-pres}}{a_{pres}} c \sim 2 \times 10^{-12} \text{ ms}^{-1} \sim 10^{-20} c . \tag{15}$$

Equally – as we determined the relations (14) and (15) – we can determine the relations for any arbitrary dimensions of Universe.

For example, according to the present cosmological literature,

*the cosmological time of Universe at the end of radiation era*

$$t_{end} \sim 3 \times 10^5 \text{ yr}. \tag{16}$$

To the relation (16) correspond

*the gauge factor of Universe at the end of radiation era*

$$a_{end} \sim 3 \times 10^{21} \text{ m} \tag{17}$$

and

*the mass of Universe at the end of radiation era*

$$m_{end} \sim 2 \times 10^{48} \text{ kg}. \tag{18}$$

It means that the mass of Universe at the end of radiation era $m_{end}$ (18) at the present cosmological time $t_{pres}$ (7) could be extended to the volume of homogeneous matter sphere with the present mass density $\rho_{pres}$ (8) and the radius (Skalský 1997)

$$r_{end-pres} = \sqrt[3]{\frac{3m_{end}}{4\pi\rho_{pres}}} \sim 3\times 10^{24}\,\text{m}. \tag{19}$$

Therefore, the matter objects which at the end of radiation era were in the distance of gauge factor $a_{end}$ (17) and, according to the relation (10), expanded at the velocity $v \sim c$ – under the hypothetical assumption of mass increase of the expansive Universe in the result of emergence of matter objects on the horizon (of the most remote visibility) – at present cosmological time $t_{pres}$ (7) would have to expand at the velocity (Skalský 1997)

$$v_{r(end-pres)} \sim \frac{r_{end-pres}}{a_{pres}}c \sim 9\times 10^{6}\,\text{ms}^{-1} \sim 3\times 10^{-2}c. \tag{20}$$

The relations (12) and (13) describe the influence of the hypothetical assumption on the emergence of matter objects on the Universe horizon (of the most remote visibility) on their expansion velocity. Concrete applications in the relations (14) and (15); and (19) and (20) clearly show that this hypothetical assumption would have to be demonstrated by rapid slowing down of matter objects expansion. However, *the theory of gravitation* (i.e. neither *the Einstein general theory of relativity*, nor *the Newton theory of gravitation (the classical mechanics)*) does not know such interaction, force, or another reason, which could cause, or explain this hypothetical slowing down of the expansion of matter objects emerging on the Universe horizon (of the most remote visibility) (Skalský 1997). The gravitation is definitely insufficient for explanation of this hypothetical slowing down of matter objects expansion, constituting the matter component of the matter-space-time structure of the Universe.

More detailed analysis of the problem of mass increase in the expansive and isotropic relativistic Universe you can see in the article: *A problem of matter increase in our observed expansive and isotropic relativistic Universe* (Skalský, 1999).

## Conclusions

The physical analysis of influence of the hypothetical assumption of the emergence of matter objects on the Universe horizon (of the most remote visibility) on their expansion velocity demonstrates unambiguously that this hypothetical assumption of the standard model of universe contradicts the Einstein general theory of relativity and its special partial solution the Newton theory of gravitation.